\documentclass[11pt,twoside]{article}
\usepackage{./asp2004}
\usepackage{psfig}
\usepackage{epsf}
\usepackage{graphics}
\usepackage{lscape}

\begin{document}
\title{On the mass distribution of neutron stars in HMXBs}
\author{A. van der Meer}
\affil{Astronomical Institute ``Anton Pannekoek'', University of
  Amsterdam, Kruislaan 403, NL-1098 SJ Amsterdam, Netherlands}
\author{L. Kaper}
\affil{Astronomical Institute ``Anton Pannekoek'', University of
  Amsterdam, Kruislaan 403, NL-1098 SJ Amsterdam, Netherlands}
\author{M.H van Kerkwijk}
\affil{Department of Astronomy and Astrophysics, University of
  Toronto, 60 St George Street, Toronto, ON M5S 3H8, Canada}
\author{E.P.J. van den Heuvel}
\affil{Astronomical Institute ``Anton Pannekoek'', University of
  Amsterdam, Kruislaan 403, NL-1098 SJ Amsterdam, Netherlands}

\begin{abstract}
  We present the results of a monitoring campaign of three eclipsing
  high-mass X-ray binaries (HMXBs: SMC~X$-$1, LMC~X$-$4 and
  Cen~X$-$3). High-resolution VLT/UVES spectra are used to
  measure the radial velocities of these systems with high
  accuracy. We show that the subsequent mass determination of the
  neutron stars in these systems is significantly improved and discuss
  the implications of this result.
\end{abstract}

\section{Introduction}

  A neutron star is the compact remnant of a massive star (M$\ga
  8$~M$_{\odot}$) with a central density of 5 to 10 times the density
  of an atomic nucleus. The global structure of a neutron star depends
  on the equation of state (EOS) under these extreme conditions,
  i.e. the relation between pressure and density in the neutron star
  interior \citep{lat04}. Given an EOS, the equations of hydrostatic
  equilibrium can be solved resulting in a mass-radius
  relation for the neutron star and a corresponding maximum
  neutron-star mass. The ``hardness'' of the EOS depends e.g. on how
  many bosons are present in matter of such a high density. As they do
  not contribute to the fermi pressure, their presence
  (e.g. \citealt{bro94}) will tend to ``soften'' the EOS. For a soft
  EOS, one of the astrophysical implications would be that neutron
  stars cannot have a large mass (e.g. $<1.55$~M$_{\odot}$ for the EOS
  applied by \citet{bro94}); for a higher mass, the object would
  collapse into a black hole.

  Given the above, the accurate measurement of neutron-star masses is
  essential for our understanding of the equation of state (EOS) of
  matter at supra-nuclear densities. Just by finding one neutron star
  with a mass higher than the maximum mass allowed by a given EOS
  proves the invalidity of that EOS. Currently, the most massive
  neutron star is the X-ray pulsar Vela~X$-$1
  \citep{bar01,qua03} with a mass of $1.86 \pm 0.16$~M$_{\odot}$. The
  millisecond radio pulsar J0751+1807 may have an even higher mass:
  $2.2 \pm 0.2$~M$_{\odot}$ \citep{sta04}. Both results are in favor
  of a stiff EOS (see also \citealt{sri01}).

  Another issue is the neutron-star mass distribution: the detailed
  supernova mechanism producing the neutron star is not understood,
  but it is likely that the many neutrinos that are produced during
  the formation of the (proto-) neutron star in the center of the
  collapsing star play an important role \citep{burr00}. \citet{tim96}
  present model calculations from which they conclude that Type~II
  supernovae (massive, single stars) give a bimodal neutron-star mass
  distribution, with peaks at 1.28 and 1.73 M$_{\odot}$, while Type~Ib
  supernovae (such as produced by stars in binaries, which are
  stripped of their envelopes) will produce neutron stars within a
  small range around 1.32~M$_{\odot}$. The massive neutron star in
  Vela~X-1 would be consistent with the second peak in this
  distribution.

  Neutron stars are detected either as radio pulsars, single or in a
  binary with a white dwarf or neutron star companion, or as X-ray
  sources in binaries with a (normal) low-mass companion star (LMXB)
  or a high-mass companion (HMXB). Most HMXBs are Be/X-ray binaries,
  but we focus on the initially more massive systems:
  high-mass X-ray binaries (HMXBs) consisting of a massive OB
  supergiant and a neutron star or a black hole (Kaper \& Van der
  Meer, these proceedings). In five of these systems containing an
  eclipsing X-ray pulsar the mass of the neutron star has been
  determined. The masses of all but one (Vela~X$-$1) are consistent
  with being equal to $1.4$~M$_\odot$. However, most spectroscopic
  observations used for these mass determinations were carried out
  more than 20 years ago, before the advent of sensitive CCD detectors
  and 8m-class telescopes; therefore, the uncertainties are too large
  to measure a significant mass difference (see \citealt{ker95b}).

\section{Observations}
  Here we present new, more accurate determinations of the
  mass of the neutron star in three of these systems, i.e. SMC~X$-$1,
  LMC~X$-$4 and Cen~X$-$3 using the high-resolution Ultraviolet and
  Visual Echelle Spectrograph UVES on the {\it Very Large Telescope}
  (VLT). These systems are in a phase of Roche-lobe overflow, have
  well determined, circular orbits ($P_{\rm orb}$ of a few days), and
  an optical counterpart of $V \simeq 14$~mag, i.e. well within reach
  of VLT/UVES.

  Cen~X$-$3 is located in our own galaxy and was identified as the
  first X-ray pulsar \citet{cho67,gia71}. The system consists of an
  O6-7 II-III star \citep{ash99} with a neutron star in a 2.09~day
  orbit \citep{nag92}. SMC~X$-$1 is located in the Small Magellanic
  Cloud (SMC). The system hosts a B0.5 Ib supergiant \citep{hut77}
  with a neutron star in a 3.89~day orbit \citep{nag92}.
  LMC~X$-$4 is located in the Large Magellanic Cloud (LMC) and
  consists of a $\sim 20$ M${_\odot}$ O8~III star (Kaper~et~al., 2005)
  and a neutron star in a 1.41~day orbit \citep{lev00}.
  \begin{figure}[!t]
  \centering
  \plotone{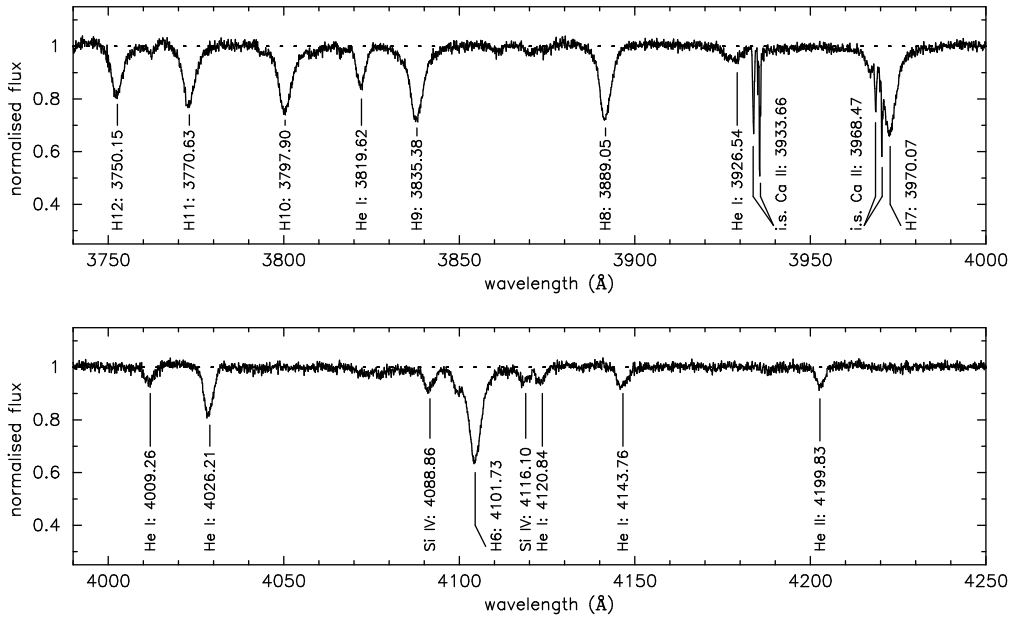}
  \caption{Spectrum of SMC~X$-$1 in the wavelength range
  3750--4250~{\AA}, obtained by the blue arm of the UVES
  spectrograph at orbital phase $\phi \sim 0.51$. Many \ion{H}{} and
  \ion{He}{} lines are detected as well as a few metal lines. The
  spectrum has a S/N $\sim 45$.}
  \label{spectrum_smcx1}
  \end{figure}

  We have obtained a dozen spectra of all three systems with VLT/UVES
  in service mode in the period October 2001 to March 2002. The
  spectra have a resolving power of $\sim 40~000$, sufficient to
  measure line positions with an accuracy of a few
  km~s$^{-1}$ and cover the wavelength range
  3600--6600~{\AA}. Figure~\ref{spectrum_smcx1} shows part of the
  spectrum of SMC~X$-$1. The spectra of LMC~X$-$4 and Cen~X$-$3 are of
  similar quality. For an extensive description of the full dataset,
  we refer to Van~der~Meer~et al. (2005).

\section{Spectral Analysis}

  To obtain a radial velocity measurement, often the complete
  spectra are cross correlated with a template spectrum. In our case
  the spectra are of such high quality that a radial velocity can be
  determined for each line separately. The advantage of such a strategy
  is that it is possible to map the influence of possible distortions
  due to X-ray heating, gravitational darkening and geometry of the
  line formation region. Here we select three lines, i.e. \ion{H}{i}
  at 3797.90~{\AA}, \ion{He}{i} at 4026.21~{\AA} and \ion{He}{i} at
  4471.50~{\AA}, that are not affected by the stellar wind and that
  can be clearly separated from other lines. In Van~der~Meer~et~al.
  (2005) we present a velocity moment analysis to select the lines
  with objective criteria. To determine the line centre we fit the
  lines with a gaussian profile. In this way we obtain a radial
  velocity curve, which we fit with a sinusoidal profile (i.e. a
  circular orbit).

\begin{figure}[!t]
  \centering
  \plottwo{vandermeer_fig2.ps}{vandermeer_fig3.ps}

  \vspace{0.5cm}
  \plottwo{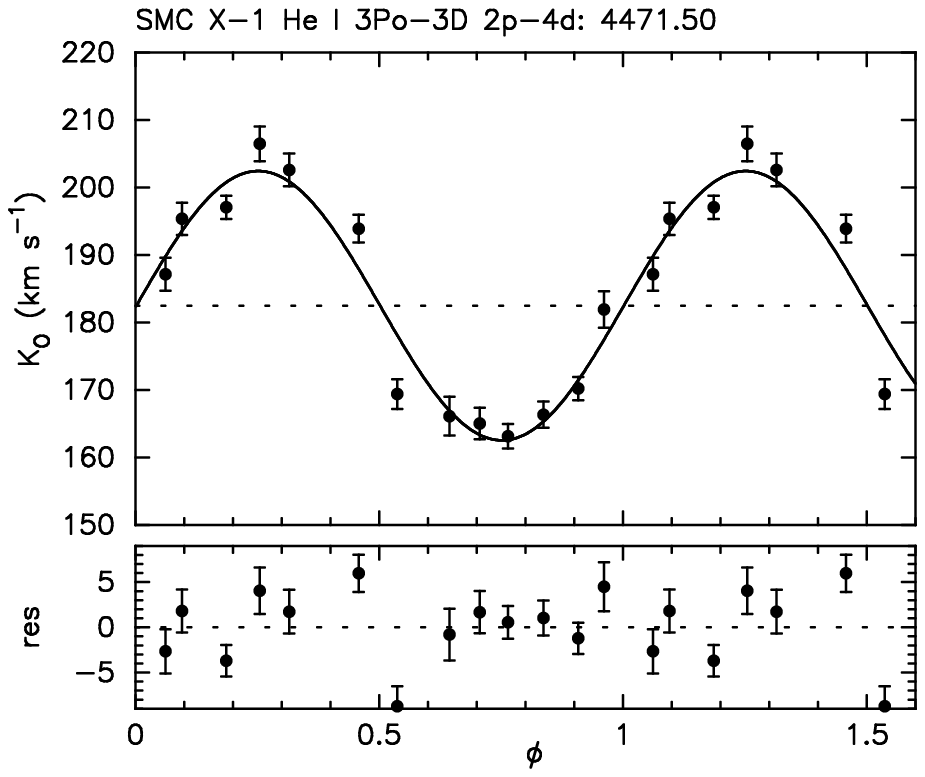}{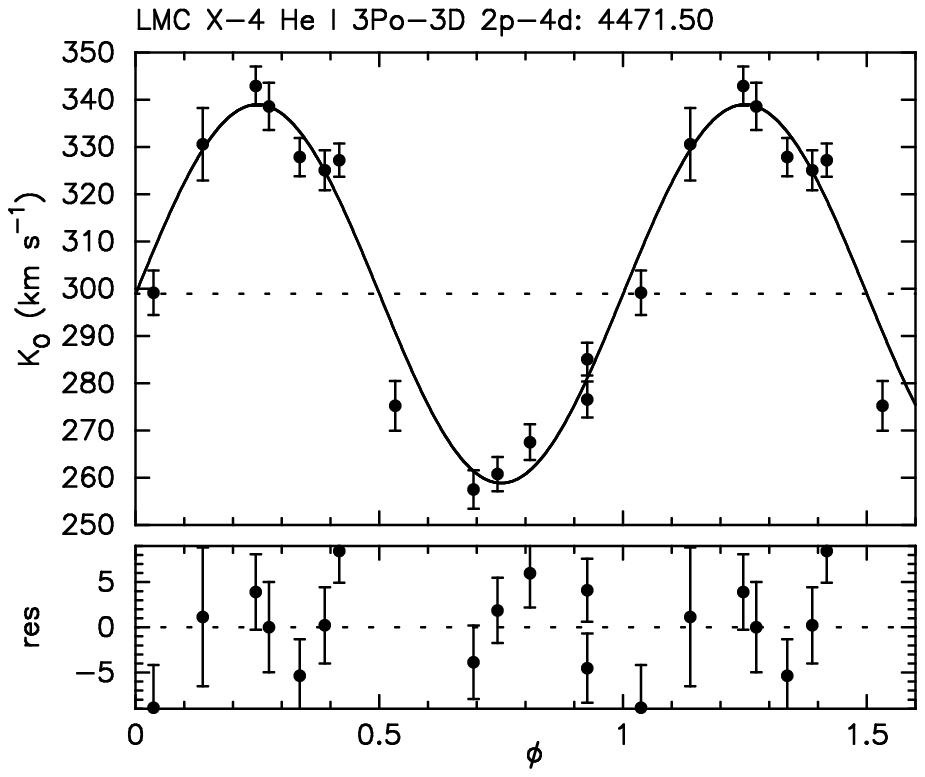}
  \caption{For each selected line that is clearly separated, that has
  a unique identification and is not affected by the stellar wind of
  the OB-supergiant a radial velocity can be determined. Here the
  velocity amplitudes of the \ion{H}{i} line at 3797.90~{\AA} and the
  \ion{He}{i} line at 4471.50~{\AA} with their fitted radial velocity
  curve is shown for SMC~X$-$1 and LMC~X$-$4. The lower panels show
  the residuals of the fit to the datapoints.}
  \label{radvels}
  \end{figure}

  The X-ray pulsar's orbital period is accurately known from pulse
  time delay measurements. To determine the orbital
  phase of the system we use the ephemeris of \citet{woj98},
  \citet{lev00} and \citet{nag92} for SMC~X$-$1, LMC~X$-$4 and
  Cen~X$-$3, respectively. The orbits are circular. The fit parameters
  that remain are the amplitude of the radial velocity curve, $K_{\rm
  O}$, and the $\gamma$~velocity of the system, $v_{\gamma}$. A few
  example radial velocity curves are shown in Fig.~\ref{radvels} for
  the \ion{H}{i} line at 3797.90~{\AA} and the \ion{He}{i} line at
  4471.50~{\AA}.

\section{Neutron Star Mass}

  In order to measure the mass of the neutron star and its optical
  companion we apply the mass function. For a circular orbit it can be
  shown that this is defined as:
  \begin{equation} 
    M_{\rm X} = \frac{K_{\rm O}^{3}P} {2\pi G \sin^{3} i} \left( 1+\frac{K_{\rm X}}{K_{\rm O}}\right)^{2},
  \end{equation}
  where $M_{\rm O}$ and $M_{\rm X}$ are the mass of the optical
  component and the X-ray source, respectively, $K_{\rm O}$ and
  $K_{\rm X}$ are the semi-amplitude of the radial velocity curve, $P$
  is the period of the orbit and $i$ is the inclination of the orbital
  plane to the line of sight.
  The value for $K_{\rm X}$ and $P$ can be obtained very accurately
  from X-ray pulse timing delay measurements. Our
  observations provide a value for $K_{\rm O}$. For the determination
  of the inclination of the system we use the approach of
  \citet{rap83}, who showed that:
  \begin{equation} 
    \sin i \approx \frac{\sqrt{1 - \left(\frac{R_{\rm L}}{a^{\prime}}\right)^{2}}} {\cos \theta_{e}} 
  \end{equation} 
  In this form the equation only holds for an optical component
  filling its Roche lobe, where $R_{\rm L}$ is its Roche-lobe radius,
  $a^{\prime}$ is the separation of the centres of masses of the two
  components, and $\theta_{\rm e}$ is the semi-eclipse angle of the
  compact object. Since in most of these systems soft X-rays are
  absorbed by the extended stellar wind of the optical companion, the
  eclipse lasts longer at low energies (see
  e.g. \citealt{hab94,mee04}). A more accurate value is obtained from
  high energy measurements. The ratio of the Roche-lobe radius and the
  orbital separation can be approximated by:
  \begin{equation} 
    \frac{R_{L}}{a^{\prime}} \approx 0.376 - 0.227 \log \frac{K_{\rm O}}{K_{\rm X}} - 0.028 \log^{2} \frac{K_{\rm O}}{K_{\rm X}}
  \end{equation} 
  The value of the constants were determined by \cite{rap83}. These
  depend on the ratio of the rotational period of the optical
  companion and the orbital period of the system. For a system undergoing
  Roche-lobe overflow, i.e. the system is tidally locked, one can
  assume that this ratio equals 1. This can be verified by determining the
  rotational period of the optical companion separately by measuring
  its (projected) rotational velocity.

  Using the formulas above we determine the mass of the neutron stars
  in our sample. The results and literature values are listed in
  Table~\ref{pars_obtained}.

\section{Conclusions}

  We show that with our VLT/UVES observations the mass measurements of
  the neutron stars in HMXBs are significantly improved. The masses
  are not all higher than 1.4 M$_{\sun}$, as is the case for
  Vela~X$-$1. The mass of SMC~X$-$1, which is $1.05 \pm 0.09$
  M$_{\sun}$, is actually the lowest neutron star mass measured so
  far. Therefore, we conclude that a mass distribution of neutron
  stars is present in HMXBs. Possible distortions by gravitational
  darkening and X-ray heating, that can influence the value of the
  radial velocity amplitude of the optical companion, are discussed in
  Van~der~Meer~et~al. (2005).

  The most accurate neutron star masses have been derived for the
  binary radio pulsars. \citet{tho99} showed that most of these
  pulsars have a mass that is consistent with a small range near
  $1.35$~M$_{\odot}$. Currently, the sample contains more systems and
  it is clear that a broader range from $1.25-1.44$~M$_{\odot}$ exists
  \citep{sta04}. \Citet{heu04} suggests that the secondary formed
  neutron star in these systems tends to be the less massive of the
  two. On the other hand, our result shows that the primary formed
  neutron star can have a low mass as well. A mass distribution of
  neutron stars in binary radio pulsars as well as in HMXBs sets an
  important constraint on the formation mechanism of neutron stars.

  \begin{table*}[!t]
  \caption[]{List of all the parameters obtained from literature. All errors are 1 $\sigma$.}
  \label{pars_obtained}
  \begin{flushleft}
  \begin{tabular}{ p{2.1cm} p{2.9cm} p{2.2cm} p{1.5cm} p{2.5cm} }
  \hline \hline
  system & P (days) & a$_{x}$ sin i (lt s) & $\rm \Theta_{\rm e}$ & K$_{\rm X}$ (km~s$^{-1}$) \\
  \hline \\
  SMC~X$-$1 & $ 3.89229090(43) $ & $ 53.4876(4) $ & $29 \pm 2 $ & $ 299.595 \pm 0.002 $ \\[0.05cm]
  LMC~X$-$4 & $ 1.40839776(26) $ & $ 26.343(16) $ & $27 \pm 3 $ & $ 407.8 \pm 0.3     $ \\[0.05cm]
  Cen~X$-$3 & $ 2.08713845(5)  $ & $ 39.56(7)   $ & $34 \pm 3 $ & $ 413.2 \pm 0.7     $ \\[0.05cm]
  \hline
  \end{tabular}
  \\[0.5cm]
  \begin{tabular}{ p{2.1cm} p{2.5cm} p{2.0cm} p{2.3cm} p{2.3cm} }
  \hline \hline
  system & K$_{\rm O}$ (km~s$^{-1}$) & i ($\deg$) & M$_{\rm O}$ (M$_{\sun}$) & M$_{\rm X}$ (M$_{\sun}$) \\
  \hline \\
  SMC~X$-$1 & $ 20.3 \pm 0.9 $ & $ 68 \pm 3  $ & $15.5 \pm 1.5$ & $1.05 \pm 0.09$ \\[0.05cm]
  LMC~X$-$4 & $ 34.3 \pm 1.2 $ & $ 65 \pm 4  $ & $15.6 \pm 1.8$ & $1.31 \pm 0.14$ \\[0.05cm]
  Cen~X$-$3 & $ 26.0 \pm 2.5 $ & $ 73 \pm 10 $ & $19.7 \pm 4.3$ & $1.24 \pm 0.24$ \\[0.05cm]
  \hline
  \end{tabular}
  \end{flushleft}
  \end{table*}

\end{document}